\documentclass[12pt]{book}

\usepackage[dvips]{graphicx,color}
\usepackage{makeidx,tsukuba}

\makeauthorindex
\makeindex

\begin{document}

\BookTitle{\itshape The 28th International Cosmic Ray Conference}
\CopyRight{\copyright 2003 by Universal Academy Press, Inc.}
\pagenumbering{arabic}

\chapter{Search for TeV Emission at the Location of Milagro Sky Survey Hot Spot Using the Whipple Gamma-Ray Telescope}

\author{%
%
%
A.~Falcone,$^{1,2}$ I.H.~Bond, P.J.~Boyle, S.M.~Bradbury,
J.H.~Buckley, D.~Carter-Lewis, O.~Celik, W.~Cui, M.~Daniel,
M.~D'Vali, I.de~la~Calle~Perez, C.~Duke, D.J.~Fegan, S.J.~Fegan,
J.P.~Finley, L.F.~Fortson, J.~Gaidos, S.~Gammell, K.~Gibbs,
G.H.~Gillanders, J.~Grube, J.~Hall, T.A.~Hall, D.~Hanna,
A.M.~Hillas, J.~Holder, D.~Horan, A.~Jarvis, M.~Jordan,
G.E.~Kenny, M.~Kertzman, D.~Kieda, J.~Kildea, J.~Knapp, K.~Kosack,
H.~Krawczynski, F.~Krennrich, M.J.~Lang, S.~LeBohec, E.~Linton,
J.~Lloyd-Evans, A.~Milovanovic, P.~Moriarty, D.~Muller, T.~Nagai,
S.~Nolan, R.A.~Ong, R.~Pallassini, D.~Petry, B.~Power-Mooney,
J.~Quinn, M.~Quinn, K.~Ragan, P.~Rebillot, P.T.~Reynolds,
H.J.~Rose, M.~Schroedter, G.~Sembroski, S.P.~Swordy, A.~Syson,
V.V.~Vassiliev, S.P.~Wakely, G.~Walker, T.C.~Weekes,
J.~Zweerink \\
{\it
(1) Purdue University, West Lafayette, IN, USA \\
(2) The VERITAS Collaboration--see S.P.Wakely's paper} ``The VERITAS
Prototype'' {\it from these proceedings for affiliations}
}

\section*{Abstract}
A recent report from the Milagro collaboration included an all-sky map created using one year of data from Milagro (peak sensitivity at 3-4 TeV for Crab-like spectra).  This map included an unidentified excess that was brighter than the Crab and was the second brightest spot on the map.  The hot spot was within the error box of the EGRET unidentified source 3EG J0520+2556.  No strong and steady emission was detected by the Whipple telescope at, or in the vicinity of, either position.  The 95\% confidence level flux upper limits from Whipple observations at the locations of the Milagro hot spot and the EGRET UnID position are 0.09 Crabs and 0.14 Crabs, respectively.

\section{Introduction}

During recent years, the field of very high energy (VHE) gamma-ray astronomy has firmly established itself by detecting and studying a handful of sources at a high significance level [1, 2].  Several of these sources have been confirmed by independent instruments and detected during many different time periods.  However, the number of known VHE sources is small.  The low number of sources may be due to the transient nature of the sources and/or the intrinsically low flux of most emitters of VHE gamma rays.  The paucity of known sources has led to many statistical limitations when attempts are made to interpret the data.  For this reason, it is of utmost importance to discover new sources and expand the number of sources and the range of source types that can be studied.

The limited number of sources has driven researchers to pursue two primary methods of increasing the VHE source catalog.  One technique has been to build instruments that can detect lower flux levels, such as HESS, MAGIC, VERITAS, and CANGAROO-III [3].  Another approach has attempted to overcome the low duty cycle of imaging air \v{C}erenkov telescopes (IACTs) by building upon extensive air shower array technology.  Examples of instruments that have pursued the latter approach are Milagro [4] and the Tibet array [5].  The disadvantage of this approach is that typical sensitivities are significantly less than those of the IACTs, while the advantage is that the entire overhead sky can be continuously monitored.

\section{Milagro and EGRET Observations}

The Milagro gamma-ray observatory, located near Los Alamos, New Mexico, is continuously scanning the entire overhead northern hemisphere.  The Milagro collaboration has reported on a survey of the northern sky using one year of data, spanning 15 Dec 2000 to 15 Dec 2001 [6].  To produce a sky map, these data were first binned into 0.1$^{\circ}$ x 0.1$^{\circ}$ bins, which is significantly smaller than the nominal angular resolution of the instrument ($\sim$0.75$^{\circ}$).  These data were then summed into an optimal bin size of 2.1 degrees [6].  The calculation of the number of independent bins on the Milagro sky map is not straightforward, and it has not been attempted here.  The skymap that is produced from this data has three hot spots.  The brightest spot is at the location of Mrk 421.  The second brightest spot on the map (RA 79.6$^{\circ}$, Dec +25.9$^{\circ}$) was not identified with any known source, but it is within the error box of the EGRET UnID object 3EG J0520+2556.  This "hot spot" produced a 4.7$\sigma$ excess (pre-trials), and it was brighter than the Crab, which was detected at 3.8$\sigma$ in this data set [6].

The EGRET UnID object 3EG J0520+2556 (RA 80.14$^{\circ}$, Dec +25.75$^{\circ}$) was detected with a significance of 6.2$\sigma$ above 100 MeV by EGRET on board the Compton Gamma Ray Observatory, but it has no known association [7].  The radius of the 95\% confidence circle has been reported to be 0.86 degrees.  The flux was measured as 15.7 +/- 2.7 x 10$^{-8}$ photons cm$^{-2}$ s$^{-1}$, and a power law spectral index of 2.83 +/- 0.24 was measured.

\section{Whipple Observations}

Between Nov. 2002 and Jan 2003, the Whipple 10 meter gamma-ray observatory pointed at the location of the Milagro hot spot and the EGRET unidentified object using two standard modes.  In both of these modes, the Crab is used to calibrate the response and the cuts of the instrument.  The peak energy response of the current configuration to a Crab-like spectrum is estimated to be $\approx$390 GeV.  The first analysis mode, referred to as TRACKING, observes only at the sky coordinates of the source and utilizes the alpha angle parameter (angle of shower image axis relative to the camera center) to perform background subtraction since gamma-ray events from the source will preferentially produce an image with a small alpha angle.  While this method allows more data to be taken at the desired location, it requires an accurate calculation of the tracking ratio, which is the ratio of events with alpha angle between 20$^\circ$-65$^\circ$ to events with alpha angle between 0$^\circ$-15$^\circ$ degrees.  By using 64 hours of OFF source data from this season, this ratio has been calculated and applied to this analysis.  The second standard analysis mode, referred to as ON/OFF, requires that the 28 minute run that is pointed at the candidate source be followed by another 28 minute run that is offset in right ascension by 30 minutes such that it moves through a blank region of sky at the same elevation and azimuth range as the original source pointing.  To perform a 2D analysis in a straightforward manner, ON/OFF data is required.  For a more detailed description of the standard analysis and the use of various cuts, refer to Mohanty et al. [8] and Reynolds et al. [9].

At the location of the Milagro hot spot, Whipple took 17 runs in TRACKING mode and 8 ON/OFF pairs.  Some data had to be discarded due to bad weather and instrumental problems, resulting in 305 minutes of TRACKING data and 139min/139min of ON/OFF data that are high quality.  Whipple also pointed directly at the location of 3EG J0520+2556 and recorded high quality data for 166 minutes in TRACKING mode and 83min/83min in ON/OFF mode.  Both pointings resulted in recorded rates that were consistent with no source at either location.  The TRACKING data taken on the location of the Milagro hot spot led to a 95\% flux upper limit of 0.09 times the flux of the Crab, using the method of Helene [10], and the TRACKING data from the location of the EGRET unidentified object led to a 95\% flux upper limit of 0.14 Crabs.

Due to the errors on the positions of both the EGRET UnID and the Milagro hot spot, a 2D analysis was also performed on the Whipple data.  The details of the two dimensional analysis are beyond the scope of this paper, however the technique is described elsewhere [11].  No significant emission was found within the 2.6 degree field of view around either source.  While there was an excess recorded near the edge of the field of view of the Milagro hot spot pointing, the post-trials significance was found to be consistent with chance.  In order to be certain of this, Whipple pointed at this excess position and recorded two ON/OFF pairs that were used to deduce a 95\% upper limit of 0.2 Crabs at this location.

\section{Discussion and Conclusions}

Based upon the Whipple data, it is clear that there is no steady VHE gamma-ray source above the 0.1 Crab flux level at the position of 3EG J0520+2556 or at the location of the Milagro hot spot.  For the Milagro excess to have been due to a steady source, rather than a statistical fluctuation or a transient source, the required flux would have been at, or in excess of, the level of the Crab.  Furthermore, the existence of any particularly strong and steady VHE sources in the immediate vicinity of these two locations has been ruled out based upon a two dimensional analysis.  However, this data can not completely dismiss the Milagro hot spot as a statistical fluctuation since it could have been a transient event from a previously undetected object such as a blazar in a high emission state or it could have been due to a source with an unusually hard spectrum.  While the latter scenario seems more unlikely than the statistical fluctuation or the transient scenarios, it will be easy to explore with additional Milagro data that has already been taken [12].  In any case, we are left with the firm conclusion that there is no steady emission (at $\sim$390 GeV assuming a Crab-like spectrum) above 0.23 Crabs from the region of either the EGRET UnID or the Milagro hot spot.

\section{Acknowledgements}

We acknowledge the technical assistance of E. Roache and J. Melnick.  This research is supported by grants from the U.S. Department of Energy, by Enterprise Ireland and by PPARC in the UK.

\section{References}

\vspace{\baselineskip}

\re
1.\ Holder J.\ et al.\ 2003, ApJL 583, L9
\re
2.\ Catanese M., Weekes T.\ 1999, Publications of the Astronomical Society of the Pacific 111, 1193
\re
3.\ Weekes T.\ 2001, proc. of Gamma-Ray Astrophysics 2001: AIP conference proceedings 587, 942
\re
4.\ Blaufuss E.\ et al.\ 2003, in preperation for Proc. of 2nd VERITAS Symposium on TeV Astrophysics of Extragalactic Sources
\re
5.\ Amenomori M.\ et al.\ 1999, ApJL 525, L93
\re
6.\ Sinnis C.\ 2002, proceedings of American Physical Society and High Energy Astrophysics of the AAS Mtg.
\re
7.\ Hartman R.\ et al.\ 1999, ApJS 123, 79
\re
8.\ Mohanty G.\ et al.\ 1998, Astroparticle Physics 9, 15
\re
9.\ Reynolds P.\ et al.\ 1993, ApJ 404, 206
\re
10.\ Helene O.\ 1983, NIM 212, 319
\re
11.\ Lessard R.\ et al.\ 2001, Astroparticle Physics 15, 1
\re
12.\ Sinnis, G.\ et al.\ 2003, these proceedings

\endofpaper
\end{document}